\documentclass{article}
\usepackage{amssymb}
\usepackage{amsfonts}
\usepackage{amsmath}

\input{tcilatex}
\begin{document}

\title{Elliptic Orbits with a Non-Newtonian Eccentricity}
\author{F.T. Hioe* and David Kuebel \and Department of Physics, St. John
Fisher College, Rochester, NY 14618, \and and \and Department of Physics \&
Astronomy, University of Rochester, Rochester, NY 14627}
\maketitle

\begin{abstract}
It is shown that the lowest order general relativistic correction produces
elliptic orbits with a non-Newtonian eccentricity.

\end{abstract}

In a weak gravitational field, the general relativistic effect of a massive
object such as a star that produces a precessional motion [1] to an
otherwise Newtonian elliptical orbit of a particle (such as a planet) is
well known and well noted for its historical significance. The precessional
angle $\Delta \phi $ is approximately given by $6\pi (GM/hc)^{2}$, where $M$
is the mass of the star, $h$ is the angular momentum per unit rest mass of
the particle, $G$ is the universal gravitation constant, and $c$ is the
speed of light. Like the precessional angle of a particle in a weak
gravitational field, for small $s\equiv GM/hc$, most lowest order general
relativistic corrections are known to be of the order of $s^{2}$ and higher.
A general relativistic correction of the order $s$, on the other hand, is
uncommon and is a principal result that we shall present in this Note.
Specifically we shall present elliptic orbits with eccentricity given by $%
\sqrt{2}s$; that is, we shall present a general relativistic effect of order 
$s$ that makes circular Newtonian orbits elliptical, and the resulting
elliptical orbits are non-precessing if terms of order $s^{2}$ and higher
can be neglected. In contrast, two new examples for hyperbolic orbits in
which the lowest order general relativistic corrections are of the order $%
s^{2}$ are also presented.

We start with one of the analytic solutions for the orbits in the
Schwarzschild geometry that we presented in our papers [2-6]. We first
introduce the parameters used in the analysis. The massive spherical object
(which we call a star) of mass $M$ sits at the origin of the coordinate
system. Let the coordinates $r$ and $\phi $ describe the position of the
particle relative to the star. If $[x^{\mu }]=(t,r,\theta ,\phi )$, then the
worldline $x^{\mu }(\tau )$, where $\tau $ is the proper time along the
path, of a particle moving in the equatorial plane $\theta =\pi /2$,
satisfies the 'combined' energy equation [1]

\begin{equation}
\overset{\cdot }{r}^{2}+\frac{h^{2}}{r^{2}}\left( 1-\frac{\alpha }{r}\right)
-\frac{c^{2}\alpha }{r}=c^{2}(\kappa ^{2}-1),
\end{equation}

where the derivative $\overset{\cdot }{}$ represents $d/d\tau $, $\alpha
\equiv 2GM/c^{2}$ is the Schwarzschild radius, $h=r^{2}\overset{\cdot }{\phi 
}$ is identified as the angular momentum per unit rest mass of the particle,
and the constant $\kappa =E/(m_{0}c^{2})$ is identified to be the total
energy per unit rest energy of the particle, $E$ being the total energy of
the particle in its orbit and $m_{0}$ the rest mass of the particle at $%
r=\infty $.

By using the dimensionless distance $q\equiv r/\alpha $ of the particle from
the star measured in units of the Schwarzschild radius and another
dimensionless quantity $U$ defined by

\begin{equation}
U\equiv \frac{1}{4}\left( \frac{\alpha }{r}-\frac{1}{3}\right) =\frac{1}{4}%
\left( \frac{1}{q}-\frac{1}{3}\right) ,
\end{equation}

eq.(1) reduces to the following simple form

\begin{equation}
\left( \frac{dU}{d\phi }\right) ^{2}=4U^{3}-g_{2}U-g_{3}
\end{equation}

where

\begin{eqnarray}
g_{2} &=&\frac{1}{12}-s^{2}  \notag \\
g_{3} &=&\frac{1}{216}+\frac{1}{6}s^{2}-\frac{1}{4}\kappa ^{2}s^{2}\equiv 
\frac{1}{216}-\frac{1}{12}s^{2}+\frac{1}{4}(1-e^{2})s^{4},
\end{eqnarray}

and where

\begin{equation}
s^{2}\equiv \left( \frac{GM}{hc}\right) ^{2},
\end{equation}

and

\begin{equation}
e^{2}\equiv 1+\frac{h^{2}c^{2}(\kappa ^{2}-1)}{(GM)^{2}}\equiv 1+\frac{%
\kappa ^{2}-1}{s^{2}}.
\end{equation}

The use of the dimensionless distance $q$ led naturally to two dimensionless
parameters $\kappa ^{2}$ (or $e^{2}$) and $s^{2}$ for characterizing the
orbit. As was pointed in our previous work [2-6], the use of the parameter $%
e^{2}$ makes the correspondence to the Newtonian case much easier to see. To
demonstrate this, we use eqs.(6) and (1) to write

\begin{equation}
e^{2}=\left( \frac{r^{2}\overset{\cdot }{\phi }}{GM}\right) ^{2}\left\{ 
\overset{\cdot }{r}^{2}+\left( r\overset{\cdot }{\phi }-\frac{GM}{r^{2}%
\overset{\cdot }{\phi }}\right) ^{2}-\frac{2GM}{c^{2}}r\overset{\cdot }{\phi 
}^{2}\right\} ,
\end{equation}

and compare this expression with the Newtonian eccentricity $e_{N}$, which,
using the derivative $\overset{\cdot }{}$ to represent $d/dt$, $t$ being the
ordinary time, can be expressed as

\begin{equation}
e_{N}^{2}=\left( \frac{r^{2}\overset{\cdot }{\phi }}{GM}\right) ^{2}\left\{ 
\overset{\cdot }{r}^{2}+\left( r\overset{\cdot }{\phi }-\frac{GM}{r^{2}%
\overset{\cdot }{\phi }}\right) ^{2}\right\} .
\end{equation}

Comparing the two above expressions, it is seen that $e\rightarrow e_{N}$ in
the Newtonian limit implies the approximation $\tau \rightarrow t$ and $%
c\rightarrow \infty $. However, since setting $c=\infty $ is not consistent
with reality, we will proceed by stating that $e\rightarrow e_{N}$ if we
take the approximation $\tau \rightarrow t$ and

\begin{equation}
\overset{\cdot }{r}^{2}+\left( r\overset{\cdot }{\phi }-\frac{GM}{r^{2}%
\overset{\cdot }{\phi }}\right) ^{2}>>\frac{2GM}{c^{2}}r\overset{\cdot }{%
\phi }^{2}.
\end{equation}

We use the coordinates $(e^{2},s^{2})$, where $-\infty \leq e^{2}\leq
+\infty $, $0\leq s\leq \infty $ of a parameter space for characterizing the
two regions which we call Regions I and II for different types of orbits
[5,6]. Region I is mathematically characterized by $\Delta \leq 0$ and
Region II is characterized by $\Delta >0$ where $\Delta $ is the
discriminant of the cubic equation

\begin{equation}
4U^{3}-g_{2}U-g_{3}=0
\end{equation}

that is defined by

\begin{equation}
\Delta =27g_{3}^{2}-g_{2}^{3}
\end{equation}

and where $g_{2}$ and $g_{3}$ are defined by eq.(4). For the case $\Delta
\leq 0$, the three roots of the cubic equation (10) are all real. We call
the three roots $e_{1},e_{2},e_{3}$ and arrange them so that $%
e_{1}>e_{2}>e_{3}$. In this paper, we are interested only in the orbit
solution for which $\Delta \leq 0$, $e_{1}>e_{2}\geq U>e_{3}$ applicable in
Region I. The equation for the orbit is [2,3]

\begin{equation}
\frac{1}{q}=\frac{1}{3}+4e_{3}+4(e_{2}-e_{3})sn^{2}(\gamma \phi ,k).
\end{equation}

The constant $\gamma $ appearing in the argument, and the modulus $k$, of
the Jacobian elliptic functions [7] are given in terms of the three roots of
the cubic equation (10) by

\begin{eqnarray}
\gamma &=&(e_{1}-e_{3})^{1/2}, \\
k^{2} &=&\frac{e_{2}-e_{3}}{e_{1}-e_{3}}.
\end{eqnarray}

where $e_{1},e_{2},e_{3}$ are given by

\begin{eqnarray}
e_{1} &=&2\left( \frac{g_{2}}{12}\right) ^{1/2}\cos \left( \frac{\theta }{3}%
\right) ,  \notag \\
e_{2} &=&2\left( \frac{g_{2}}{12}\right) ^{1/2}\cos \left( \frac{\theta }{3}+%
\frac{4\pi }{3}\right) ,  \notag \\
e_{3} &=&2\left( \frac{g_{2}}{12}\right) ^{1/2}\cos \left( \frac{\theta }{3}+%
\frac{2\pi }{3}\right) ,
\end{eqnarray}

and where

\begin{equation}
\cos \theta =g_{3}\left( \frac{27}{g_{2}^{3}}\right) ^{1/2}.
\end{equation}

A typical orbit given by eq.(12) (not on any one of the three boundaries) in
Region I is a precessional elliptic-type orbit for $e^{2}<1$, a
parabolic-type orbit for $e^{2}=1$, and a hyperbolic-type orbit for $e^{2}>1$
[5,6].

For the elliptic-type orbits ($e^{2}<1$), the maximum distance $r_{\max }$
(the aphelion) of the particle from the star and the minimum distance $%
r_{\min }$ (the perihelion) of the particle from the star, or their
corresponding dimensionless forms $q_{\max }$ $(=r_{\max }/\alpha )$ and $%
q_{\min }$ $(=r_{\min }/\alpha ),$ are obtained from eq.(12) when $\gamma
\phi =0$ and when $\gamma \phi =K(k)$ respectively, where $K(k)$ is the
complete elliptic integral of the first kind [7], and they are given by

\begin{equation}
\frac{1}{q_{\max }}=\frac{1}{3}+4e_{3},
\end{equation}

and

\begin{equation}
\frac{1}{q_{\min }}=\frac{1}{3}+4e_{2}.
\end{equation}

The geometric eccentricity $\varepsilon $ of the orbit is defined in the
range $0\leq \varepsilon \leq 1$ by

\begin{equation}
\varepsilon \equiv \frac{r_{\max }-r_{\min }}{r_{\max }+r_{\min }}=\frac{%
q_{\max }-q_{\min }}{q_{\max }+q_{\min }}=\frac{e_{2}-e_{3}}{1/6+e_{2}+e_{3}}%
,
\end{equation}

using $q_{\max }$ and $q_{\min }$ given by eqs.(17) and (18). It has been
shown in ref.3 that in the range $0\leq \varepsilon <1$ that $\varepsilon
\rightarrow e$ from below as $s\rightarrow 0$, and that $\varepsilon =e$ for
all values of $s$ when $\varepsilon =1$. 

The precessional angle $\Delta \phi $ is given by

\begin{equation}
\Delta \phi =\frac{2K(k)}{\gamma }-2\pi .
\end{equation}

The Newtonian correspondence is approached by making $s$ very small.
Substituting eq.(4) into eq.(16) and expanding in power series in $s$, we
find

\begin{equation}
\cos \theta =1-2\cdot 3^{3}e^{2}s^{4}-2^{2}\cdot 3^{3}(1+9e^{2})s^{6}-2\cdot
3^{5}\cdot 5(1+6e^{2})s^{8}+...
\end{equation}

To obtain a power series in $s$ for $\theta $, the point $e^{2}=0$ must be
done separately.

For $e^{2}>0$, we have

\begin{equation}
\theta =2\cdot 3\sqrt{3}es^{2}\left[ 1+e^{-2}(1+9e^{2})s^{2}+...\right] .
\end{equation}

Expanding $e_{1},e_{2},e_{3},\gamma ,k^{2},sn(\gamma \phi ,k)$ in eq.(12), $%
\varepsilon $ in eq.(19), and $\Delta \phi $ in eq.(20) in the power series
in $s$, we find that the orbit equation (12) can be approximated for $%
e^{2}>0 $ and for small $s^{2}$ by

\begin{equation}
\frac{1}{q}=2s^{2}\{1-\varepsilon \cos [(1-\delta )\phi ]\},
\end{equation}

which, in terms of $r$, gives the approximate orbit equation

\begin{equation}
\frac{1}{r}=\frac{GM}{h^{2}}\{1-\varepsilon \cos [(1-\delta )\phi ]\},
\end{equation}

where $\varepsilon $, to the order of $s^{2}$, is given by

\begin{equation}
\varepsilon \simeq e+(e^{-1}-e^{3})s^{2},
\end{equation}

and where $\delta $, to the order of $s^{2}$, is given by%
\begin{equation}
\delta \simeq 3s^{2}.
\end{equation}

$\delta $ is related to the precessional angle $\Delta \phi $ given in
eq.(20) by $\Delta \phi \simeq 2\pi \delta \simeq 6\pi s^{2}=6\pi \lbrack
GM/(hc)]^{2}$ and it is independent of $e$ (to the order $s^{2}$) for $%
0<e\leq \infty $. As an example, a general relativistic elliptic-type orbit
with $e=0.8$, $s=0.0176539$ has an exact $\varepsilon =0.80015$. The lowest
order general relativistic corrections (25) and (26) yield $\varepsilon
\simeq 0.80023$ and $\delta \simeq 0.000935$. The corrections to $%
\varepsilon \simeq e$ and the magnitude of $\delta $ are both of the order $%
s^{2}$.

Thus if we can ignore terms of order $s^{2}$ and higher, we recover the
Newtonian orbit equation given by

\begin{equation}
\frac{1}{r}=\frac{GM}{h^{2}}(1-e\cos \phi ).
\end{equation}

The question that can be posed at this point is whether we can define the
Newtonian limit by stating that it is the general relativistic result for
small $s$ if we ignore terms of order $s^{2}$ and higher.

To proceed, we note that the case $e=0$ is excluded from the expansion given
by eq.(22) and it is also clear that $e^{2}=0$ does not satisfy the
condition given by eq.(9) and must be treated separately.

For $e^{2}=0$, the expansion for $\theta $, instead of eq.(22), is now

\begin{equation}
\theta =\sqrt{2^{3}\cdot 3^{3}}s^{3}\left( 1+\frac{3^{2}\cdot 5}{2^{2}}%
s^{2}+...\right)
\end{equation}

and the approximate orbit equation still has the form of eq.(23) or (24)
with the same $\delta $ given by eq.(26) to the order of $s^{2}$, but with $%
\varepsilon $, instead of eq.(25), now given to the order of $s^{3}$ by

\begin{equation}
\varepsilon =\sqrt{2}s\left( 1+\frac{9}{4}s^{2}\right) .
\end{equation}

Thus we have elliptic orbits that precess with the same angle $\delta $
given by eq.(26) but with an eccentricity equal to $\sqrt{2}s+9\sqrt{2}%
s^{3}/4$ to the order $s^{3}$. If we ignore terms of order $s^{2}$ and
higher, the orbit equation becomes

\begin{equation}
\frac{1}{r}=\frac{GM}{h^{2}}(1-\sqrt{2}s\cos \phi ),
\end{equation}

which is a (non-precessing) elliptical orbit with a non-Newtonian
eccentricity that is dependent on the speed of light. To the best of our
knowledge, this general relativistic elliptical orbit with an eccentricity
of the order $s$ is new and has never been noted by other authors. As an
example, an elliptic orbit with eccentricity $\varepsilon =0.0368035$ could
come from a general relativistic orbit with $e=0$ [remembering that $e$ is
defined by eq.(7) and not eq.(8)] and $s=0.0259843$ for which the
approximation formula (29) gives $\varepsilon \simeq 0.0368032$ [the first
term alone gives $\sqrt{2}s=0.0367474$] and eq.(26) gives $\delta \simeq
0.0020256$. The approximation $\varepsilon \simeq e=0$ holds if we ignore
terms of order $s$. The orbit for $e=0$ from general relativity becomes a
Newtonian circular orbit if we ignore terms of order $s$, i.e. ignoring the
second order terms in $s^{2}$ and higher order terms is not sufficient to
get the Newtonian limit for this case.

It is clear that the entire region characterized by $e^{2}\leq 0$ or

\begin{equation}
\overset{\cdot }{r}^{2}+\left( r\overset{\cdot }{\phi }-\frac{GM}{r^{2}%
\overset{\cdot }{\phi }}\right) ^{2}\leq \frac{2GM}{c^{2}}r\overset{\cdot }{%
\phi }^{2}
\end{equation}

is non-Newtonian in character. This includes all circular orbits that occur
[6] on the curve $s^{2}=s_{1}^{\prime 2}$ for which $k^{2}=0$ and $%
\varepsilon =0$, where $s_{1}^{\prime 2}$ is given by

\begin{equation}
s_{1}^{\prime 2}=\frac{1-9e^{2}-\sqrt{(1+3e^{2})^{3}}}{27(1-e^{2})^{2}}
\end{equation}

from the "vertex" $V$ at $(e^{2},s^{2})=(-1/3,1/12)$ where the innermost
stable circular orbit (ISCO) occurs, to the origin $O$ at $%
(e^{2},s^{2})=(0,0)$ where the circular orbit has an infinite radius. This
curve $s_{1}^{\prime }$ defines a boundary of Region I for which the values
of $e^{2}$ range between $-1/3$ and $0$ and the values for $s^{2}$ range
between $1/12$ and $0$. All circular orbits precess even though the
precession angle is not observable [8], and for small $s$ the precession
angle is given by

\begin{equation*}
\Delta \phi \simeq 6\pi s^{2}\simeq 6\pi \frac{GM}{c^{2}r_{c}},
\end{equation*}

where $r_{c}\simeq h^{2}/(GM)$ is the radius of the circular orbit, and $%
\Delta \phi $ is non-zero unless the radius of the circle is infinite which
occurs on $s=0$ for zero gravitational field.

The values of $e^{2}$ along the $s_{1}^{\prime }$ curve where the circular
orbits occur near $s=0$ are given by

\begin{equation*}
e^{2}\simeq -2s_{1}^{\prime 2}.
\end{equation*}

For small $s_{1}^{\prime }$ and for $s$ just above $s_{1}^{\prime }$ inside
Region I given by $s^{2}=s_{1}^{\prime 2}+(\Delta s)^{2}$, it can be shown
in the same manner that to the order $\Delta s$ we have elliptic orbits
similar to eq.(30) given by

\begin{equation}
\frac{1}{r}=\frac{GM}{h^{2}}[1-\sqrt{2}(\Delta s)\cos \phi ].
\end{equation}

For the parabolic-type orbit ($e^{2}=1$), $e_{3}=-1/12$ and the initial
distance of the particle from the star is given from eq.(17) to be $q_{\max
}=\infty $ and eq.(19) gives $\varepsilon =1$. Thus $e=1$ and $\varepsilon
=1 $ coincide for all values of $s$. The orbit equation is given exactly by

\begin{equation}
\frac{1}{q}=4(e_{2}+\frac{1}{12})sn^{2}(\gamma \phi ,k),
\end{equation}

and for small $s$ values approximately by

\begin{equation}
\frac{1}{r}=\frac{GM}{h^{2}}\{1-\cos [(1-\delta )\phi ]\},
\end{equation}

where $\delta $ is given by eq.(26) and for which the lowest order general
relativistic correction to the Newtonian case is of the order $s^{2}$.

For the hyperbolic-type orbit ($e^{2}>1$), $e_{3}$ is less than $-1/12$ and
eq.(17) is not applicable. Instead, a particle approaches the star from
infinity along an incoming asymptote at an angle $\Psi _{1}$ to the
horizontal axis given by [2,3]

\begin{equation}
\Psi _{1}=\gamma ^{-1}sn^{-1}\left( \sqrt{-\frac{\frac{1}{3}+4e_{3}}{%
4(e_{2}-e_{3})}},k\right) ,
\end{equation}

where $\gamma $ and $k$ are defined by eqs.(13) and (14), turns
counter-clockwise about the star to its right on the horizontal axis, and
leaves along an outgoing asymptote at an angle $\Psi _{2}$ given by

\begin{equation}
\Psi _{2}\equiv \frac{2K(k)}{\gamma }-\Psi _{1}.
\end{equation}

The minimum dimensionless distance $q_{\min }$ of the particle from the star
is still given by eq.(18) as $e_{2}$ is still greater than $-1/12$.

In the Newtonian limit for small $s$, $\Psi _{1}$ becomes

\begin{equation}
\Psi _{1}\simeq \cos ^{-1}(\frac{1}{e})\equiv \phi _{0},
\end{equation}

and the complementary angle $\Psi _{1}^{\prime }\equiv 2\pi -\Psi _{2}\simeq
\phi _{0}$ also. If we define

\begin{equation}
\Theta _{GR}\equiv \Psi _{1}+\Psi _{1}^{\prime }
\end{equation}

and

\begin{equation}
\Theta _{Newton}=2\phi _{0},
\end{equation}

the difference $\Delta \phi \equiv \Theta _{Newton}-\Theta _{GR}$ can be
taken to be an analog of the precession angle given by eq.(20) for a
hyperbolic orbit, and for small $s$ and to the order of $s^{2}$, it was
shown [3] to be given by

\begin{equation}
\Delta \phi \simeq \left[ 6\pi -6\phi _{0}+2(2+e^{-2})\sqrt{e^{2}-1}\right]
s^{2}.
\end{equation}

This result is different from the approximation used by Longuski et al. [9].
As discussed in ref.9, an experimental test can be carried out to check this
result.

From eq.(18), the minimum distance $r_{\min }$ of the particle from the star
is given approximately by

\begin{equation}
\frac{1}{r_{\min }}\simeq \frac{GM}{h^{2}}\left( e+1\right) \left[ 1+\frac{%
(e+1)^{2}}{e}s^{2}+...\right] .
\end{equation}

Equations (41) and (42) show two examples for which the lowest general
relativistic corrections to the Newtonian case are of the order $s^{2}$ and
not $s$ for $e^{2}>1$.

To summarize the above results, for small values of $s$, the general
relativistic correction to the Newtonian elliptic orbit is second order in $%
s $ for $e^{2}>0$ but is first order in $s$ for $e^{2}\leq 0$ for which the
correction is shown to appear in the eccentricity of the elliptic orbit.
This division gives a new meaning and significance to the parameter $e^{2}$.
In particular, there exist non-Newtonian elliptic orbits of eccentricity $%
\sqrt{2}s$ given by eq.(30).

\bigskip

\bigskip

\bigskip

References

*Electronic address: fhioe@sjfc.edu

[1] M.P. Hobson, G. Efstathiou and A.N. Lasenby: General Relativity,
Cambridge University Press, 2006, Chapters 9 and 10.

[2] F.T. Hioe, Phys. Lett. A 373, 1506 (2009).

[3] F.T. Hioe and D. Kuebel, Phys. Rev. D 81, 084017 (2010).

[4] F.T. Hioe and D. Kuebel, arXiv:1008.1964 v1 (2010).

[5] F.T. Hioe and D. Kuebel, arXiv:1010.0996 v2 (2010).

[6] F.T. Hioe and D. Kuebel, arXiv:1207.7041v1 (2012).

[7] P.F. Byrd and M.D. Friedman: Handbook of Elliptic Integrals for
Engineers and Scientists, 2nd Edition, Springer-Verlag, New York, 1971.

[8] J.L. Martin: General Relativity, Revised Edition, Prentice Hall, New
York 1996, Chapter 4.

[9] J.M. Longuski, E. Fischbach, and D.J. Scheeres, Phys. Rev. Lett. 86,
2942 (2001), J.M. Longuski, E. Fischbach, D.J. Scheeres, G. Giampierri, and
R.S. Park, Phys. Rev. D 69, 042001 (2004).

\end{document}